\documentclass[runningheads]{llncs}
\usepackage{cite}
\usepackage{amsmath,amssymb,amsfonts}
\usepackage{algorithmic}
\usepackage{graphicx}
\usepackage{textcomp}
\usepackage{caption}
\usepackage{subfigure}
\usepackage{epstopdf}

\def\BibTeX{{\rm B\kern-.05em{\sc i\kern-.025em b}\kern-.08em
    T\kern-.1667em\lower.7ex\hbox{E}\kern-.125emX}}
\begin{document}

\title{A novel quantum visual secret sharing scheme}
\titlerunning{}
\author{{Wenjie Liu}\inst{1,2},\and {Yinsong Xu}\inst{2},\and{Maojun Zhang}\inst{3},\and{ Junxiu Chen}\inst{2},\and {Ching-Nung Yang}\inst{4}}

\institute{Jiangsu Engineering Center of Network Monitoring, Nanjing University of Information Science and Technology, Nanjing 210044, P. R. China \inst{1}\\
School of Computer and Software, Nanjing University of Information Science and Technology, Nanjing 210044, P. R. China\inst{2}\\
School of Mathematics and Computing Science, Guilin University of Electronic Technology, Guilin 541004, P. R. China\inst{3}\\
Division Department of Computer Science and Information Engineering, National Dong Hwa University, Taiwan 97441, P. R. China\inst{4}\\Corresponding author: Maojun Zhang \email{ zhang1977108@sina.com}.}

\maketitle

\begin{abstract}
Inspired by Naor \textit{et al.}'s visual secret sharing (VSS) scheme, a novel $n$ out of $n$ quantum visual secret sharing (QVSS) scheme is proposed, which consists of two phases: sharing process and recovering process. In the first process, the color information of each pixel from the original secret image is encoded into an $n$-qubit superposition state by using the strategy of quantum expansion instead of classical pixel expansion, and then these $n$ qubits are distributed as shares to $n$ participants, respectively. During the recovering process, all participants cooperate to collect these $n$ shares of each pixel together, then perform the corresponding measurement on them, and execute the $n$-qubit \textit{XOR} operation to recover each pixel of the secret image. The proposed scheme has the advantage of single-pixel parallel processing that is not available in the existing analogous quantum schemes, and perfectly solves the problem that in the classic VSS schemes the recovered image has the loss in resolution. Moreover, its experiment implementation with IBM Q is conducted to demonstrate the practical feasibility.
\end{abstract}


\section{Introduction}
\label{sec:introduction}
In order to prevent the secret from being too concentrated and achieve the purpose of spreading risk and tolerating intrusion, the idea of secret sharing \cite{Shamir79,Blakley79} has been proposed. Secret sharing refers to methods for distributing a secret amongst a group of participants, each of whom is distributed a share of the secret. The secret can be reconstructed only when a sufficient number of shares are combined together, but individual shares are of no use on their own. This method provides an effective way for the security protection and fair use of secret keys.  As an important branch of modern cryptography, secret sharing plays an important role in the preservation, transmission and utilization of data.

Inspired by secret sharing, many experts and scholars have devoted themselves to explore the research of visual cryptography, i.e., how to utilize the idea of secret sharing to solve the problem of image encryption and decryption. In 1995, the first visual secret sharing (VSS) scheme based on pixel expansion was proposed by Naor \textit{et al.} \cite{Naor95}, where a binary secret image is shared to generate a plurality of noise-like shared images by dividing each of the secret pixels into pixel blocks composed of $m$ sub-pixels, and the shared image is $m$ times the secret image. Due to the presence of pixel expansion, the visual quality of the recovered secret image may be not ideal. Since then, most of the research has been carried out around reducing pixel expansion \cite{Cimato05,Yang07,Yang04,Hsu06,Lee14}.

With the continuous development of quantum computing technology, many researchers try to apply quantum mechanics in many fields, such as quantum key distribution (QKD) \cite{Bennett84,Ekert91}, quantum key agreement (QKA) \cite{Huang17,Liu18Xu}, quantum secret sharing (QSS) \cite{Hillery99,Cleve99}, quantum secure direct communication (QSDC) \cite{Liu09,Zhong18}, quantum remote state preparation (QRSP) \cite{Liu15,Qu17}, quantum steganography (QS) \cite{Wu18,Qu18Zhu,Qu18Cheng},  delegating quantum computation (DQC) \cite{Liu17,Liu18Chen}, quantum private query (QPQ) \cite{Gao19,Liu19Gao} and even quantum machine learning \cite{Lloyd13,Liu18Gao,Liu19}. Among them, QSS is an important research area \cite{Hillery99,Cleve99,Li09,Sun10,Wei15,Qin16,Zhang05Li,Zhang05Man}, and it can be viewed as the generalization of classical secret sharing to the setting of quantum information. In 1999, Hillery \textit{et al.} proposed the first QSS scheme by using the Greenberger-Horne-Zeilinger (GHZ) state. In the scheme, a
GHZ triplet is split and each of the other two participants gets a particle. Both participants are allowed to measure their particles in either $x$ or $y$ direction, and their results are combined to determine the dealer's measurement result. This allows dealer to establish a joint key with two participants which dealer can then use to send message. At the same year, a threshold QSS scheme \cite{Cleve99} was proposed by adopting quantum error correcting code theory. Since then, various kinds of QSS schemes are constantly being proposed \cite{Li09,Sun10,Wei15,Qin16,Zhang05Li,Zhang05Man}. All these work can be divided into three main categories, QSS of classical messages \cite{Li09,Sun10}, QSS of quantum information \cite{Wei15,Qin16} where the secret is an arbitrary unknown state, and QSS of both them \cite{Zhang05Li,Zhang05Man}.

As far as we know, the literatures on how to use quantum mechanisms to solve VSS schemes are rare \cite{Song14,Das15}. In 2014, Song \textit{et al.} proposed a flexible (${2^k}$, ${2^k}$) quantum image secret sharing (QISS) scheme \cite{Song14}, where the whole secret image is repeatedly encoded into a quantum state, and then split into sub-images as shares with multiple measurement operations. Although the size of each share (the part of the original quantum image) becomes smaller, it requires a mass of multi-qubit superposition states to produce shares by measurement, and also loses the characteristic of single-pixel parallel processing (i.e., one pixel as a unit for parallel processing) in VSS. In order to solve the problem, we propose a novel $n$ out of $n$ quantum visual secret sharing (QVSS) scheme based on Naor \textit{et al.}'s scheme. In our scheme, the color information of each pixel from the original secret image is encoded into an $n$-qubit quantum superposition state, so the advantage of single-pixel parallel processing can be preserved. Besides, we use the quantum expansion strategy to perfectly solve the problem that the recovered image has the loss in resolution in the classical VSS schemes, i.e., the recovered image is as same as the original secret image.

The rest of this paper is organized as follows. Sect. \ref{sec:2} provides some preliminary knowledge about quantum computation and Naor \textit{et al.}'s VSS scheme. The proposed ($n$, $n$) QVSS scheme, which consists of sharing process and recovering process, is explicated in Sect. \ref{sec:3}, and an example, ($3$, $3$) QVSS scheme, is illustrated in Sect. \ref{sec:4}. And then, the correctness is verified in Sect. \ref{sec:5}. We also compare and discuss with other analogous schemes in Sect. \ref{sec:6}. Moreover, its experiment implementation with IBM Q is conducted to demonstrate the practical feasibility in Sect. \ref{sec:7}. Finally, Sect. \ref{sec:8} gives the discussion and conclusion of this paper.

\section{Preliminaries}
\label{sec:2}
\subsection{Quantum computation}
As we know, the bit is the fundamental concept of classical information, and has a state, either 0 or 1. Similar to the classical bit, the quantum bit (called qubit) \cite{Nielsen02} is the basic unit of quantum information and has two possible states $\left| 0 \right\rangle $ and $\left| 1 \right\rangle $, which is often referred to as quantum superposition state,

\begin{equation}\label{eqn1}
  \left| \varphi  \right\rangle  = \alpha \left| 0 \right\rangle  + \beta \left| 1 \right\rangle,
\end{equation}
where $\alpha $, $\beta $ are complex numbers, and ${\left| \alpha  \right|^2} + {\left| \beta  \right|^2} = 1$. $\left| 0 \right\rangle $ and $\left| 1 \right\rangle $ can be represented by vectors,

\begin{equation}
  \left| 0 \right\rangle  = \left[ \begin{array}{l}
1\\
0
\end{array} \right],  {\kern 1pt} {\kern 1pt} {\kern 1pt} {\kern 1pt} {\kern 1pt} {\kern 1pt} {\kern 1pt} {\kern 1pt} {\kern 1pt} {\kern 1pt} {\kern 1pt} {\kern 1pt} {\kern 1pt} {\kern 1pt} {\kern 1pt} \left| 1 \right\rangle  = \left[ \begin{array}{l}
0\\
1
\end{array} \right].
\label{eqn2}
\end{equation}
Then, $\left| \varphi  \right\rangle$ can be expressed in vector form $\left| \varphi  \right\rangle  = \left( \begin{smallmatrix}
\alpha \\
\beta
\end{smallmatrix} \right)$.

Analogous to the way that a classical computer is built from an electrical circuit containing wires and logic gates, a quantum computer is built from a quantum circuit containing wires and elementary quantum gates to carry around and manipulate the quantum information. Single-qubit gates, such as \textit{Pauli-X}, \textit{Pauli-Z}, and \textit{H} (\textit{Hadamard}), are the simplest form of quantum gates, and they can be described as $2 \times 2$ unitary matrices as below,

\begin{equation}
  \textit{X} = \left[ {\begin{matrix}
0 & 1  \\
1 & 0
\end{matrix}} \right],{\kern 4pt} \textit{Z} = \left[ {\begin{matrix}
 1&0 \\
  0&-1 \\
\end{matrix}} \right],{\kern 4pt}
\textit{H}  =  \frac{1}{{\sqrt 2 }}\left[ {\begin{matrix}
1&1\\
1&{ - 1}
\end{matrix}} \right].
\label{eqn3}
\end{equation}

\begin{figure}[htbp]
  \centering
  \includegraphics[width=3in]{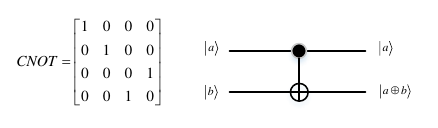}\\
  \caption{Matrix representation and quantum circuit of \textit{CNOT} gate.}\label{fig1}
\end{figure}

Multi-qubit gates are also the important units in a quantum circuit. The prototypical multi-qubit quantum logic gate is \textit{controlled-NOT} (i.e., \textit{CNOT}) gate (shown in Fig. \ref{fig1}), which has two input qubits, known as the control qubit and the target qubit, respectively. If the control qubit is set to 0, then the target qubit is left alone. If the control qubit is set to 1, then the target qubit is flipped.

\begin{figure}[htbp]
  \centering
  \includegraphics[width=2.5in]{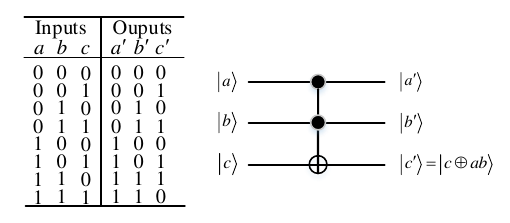}\\
  \caption{Truth table and quantum circuit of \textit{Toffoli} gate.}\label{fig2}
\end{figure}

Besides \textit{CNOT} gate, \textit{Toffoli} gate is another frequently used multi-qubit gate. As illustrated in Fig. \ref{fig2}, \textit{Toffoli} gate has three input bits and three output bits: two of the bits are control bits that are unaffected by the action of the \textit{Toffoli} gate; the third bit is a target bit that is flipped if both control bits are set to 1, and otherwise is left alone.

\subsection{Naor et al.'s Visual secret sharing scheme}
\begin{figure*}[htbp]
  \centering
  \includegraphics[width=6in]{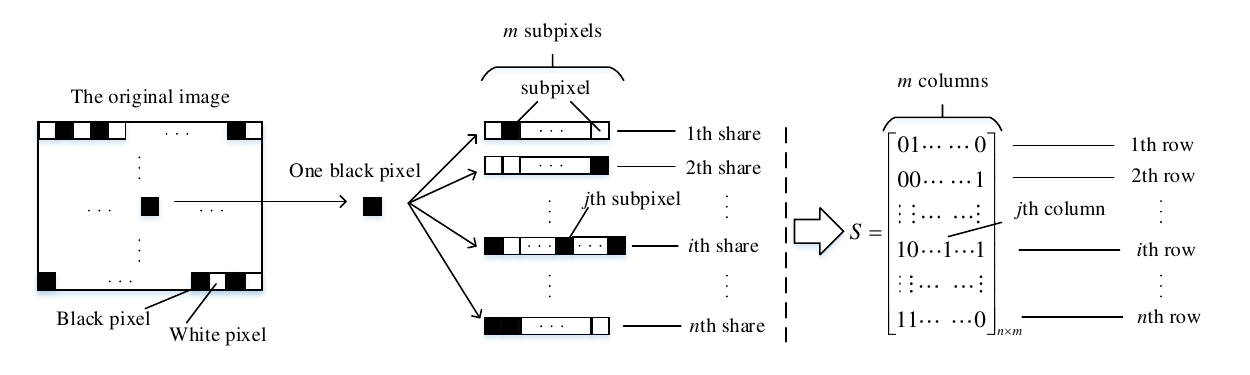}\\
  \caption{The pixel expansion process in Naor \textit{et al.}'s VSS scheme}\label{fig3}
\end{figure*}

The first visual secret sharing scheme was proposed by Naor \textit{et al.} \cite{Naor95}, where the secret image consists of a collection of black and white pixels. As shown in Fig. \ref{fig3}, each original pixel generates $n$ shares, and each share has a collection of $m$ black and white subpixels (the process is named pixel expansion). These $n \times m$ subpixels can be described by an $n \times m$ Boolean matrix $S = [{s_{ij}}]$, where $i \in \{ 1,2, \cdots ,n\} $, $j \in \{ 1,2, \cdots ,m\} $, and ${s_{ij}} \in \{ 0,1\} $. If ${s_{ij}} = 1$, the $j$th subpixel in the $i$th share is black, otherwise, it white.

In the ($k$, $n$) VSS scheme, it mainly consists of two matrix sets ${C_0}$ and ${C_1}$ which are composed of $n \times m$ Boolean matrices $S$. To share a white pixel, the dealer randomly chooses one $S$ in ${C_0}$, and to share a black pixel, the dealer randomly chooses one $S$ in ${C_1}$. The chosen matrix $S$ defines the colour of the $m$ subpixels in each one of the $n$ shares. The scheme is considered valid if the following two conditions are met:

(1) $\forall {\kern 1pt} S{\kern 1pt}  \in {\kern 1pt} {\kern 1pt} {C_0}$, $\left\{ {{i_1},{i_2}, \cdots ,{i_p}} \right\} \subseteq \left\{ {1,2, \cdots ,n} \right\}\left( {p \geqslant k} \right)$, vector $V = S[{i_1}] + S[{i_2}] +  \cdots  + S[{i_k}]$, so $H(V) \leqslant d - \alpha m$, where "$+$" means "\textit{OR}" logical operation, $S[i]$ indicates the $i$th row of $S$, $d$ and $\alpha $ are threshold and relative difference respectively, and $H(V)$ represents the Hamming weight of $V$. $\forall {\kern 1pt} S{\kern 1pt}  \in {\kern 1pt} {\kern 1pt} {C_1}$, $H(V) \geqslant d$.

(2) $\forall \left\{ {{i_1},{i_2}, \cdots ,{i_q}} \right\} \subseteq \left\{ {1,2, \cdots ,n} \right\}\left( {q < k} \right)$, the two collections of $q \times m$ matrices ${D_t}$ for $t \in \{ 0,1\} $ obtained by restricting each $n \times m$ matrix in ${C_t}$ ($t \in \{ 0,1\} $) to rows ${i_1},{i_2}, \cdots ,{i_q}$ are indistinguishable in the sense that they contain the same matrices with the same frequencies.

When the secret image needs to be recovered, any $k$ participants just print their shares on transparencies respectively and stack their transparencies to decrypt secret information by the human visual system. But any $k-1$ of them gain no information about secret information.

As mentioned above, the pixel is treated as a separate basic unit throughout the scheme, so this single-pixel based VSS scheme is well suitable for parallel computing of images. In order to make individual participant get no information, most VSS schemes use the strategy of pixel expansion to confuse the information of subpixels. However, it may cause the loss in resolution from the original image to the recovered one, i.e., the size of the shared image is $m$ times the original secret image. Therefore, some researchers tried to reduce pixel expansion \cite{Cimato05,Yang07}, and even implement non-expansion \cite{Yang04,Hsu06,Lee14}.

\section{A novel ($n$, $n$) quantum visual secret sharing scheme}
\label{sec:3}
Suppose the dealer Alice wants to share her secret image, which consists of $s$ black or white pixels, to $n$ participants $\left\{ {Bo{b_1}, {\kern 1pt} Bo{b_2}{\kern 1pt}, \cdots , {\kern 1pt} Bo{b_n}} \right\}$, and the black and white pixels are respectively represented by  $|1\rangle $ and $|0\rangle $. The secret image can be recovered by $n$ participants, but it cannot be less than $n$ participants. The specific ($n$, $n$) QVSS scheme is mainly composed of the sharing process and the recovering process.

\subsection{Sharing process}
\begin{figure*}[htbp]
  \centering
  \includegraphics[width=6in]{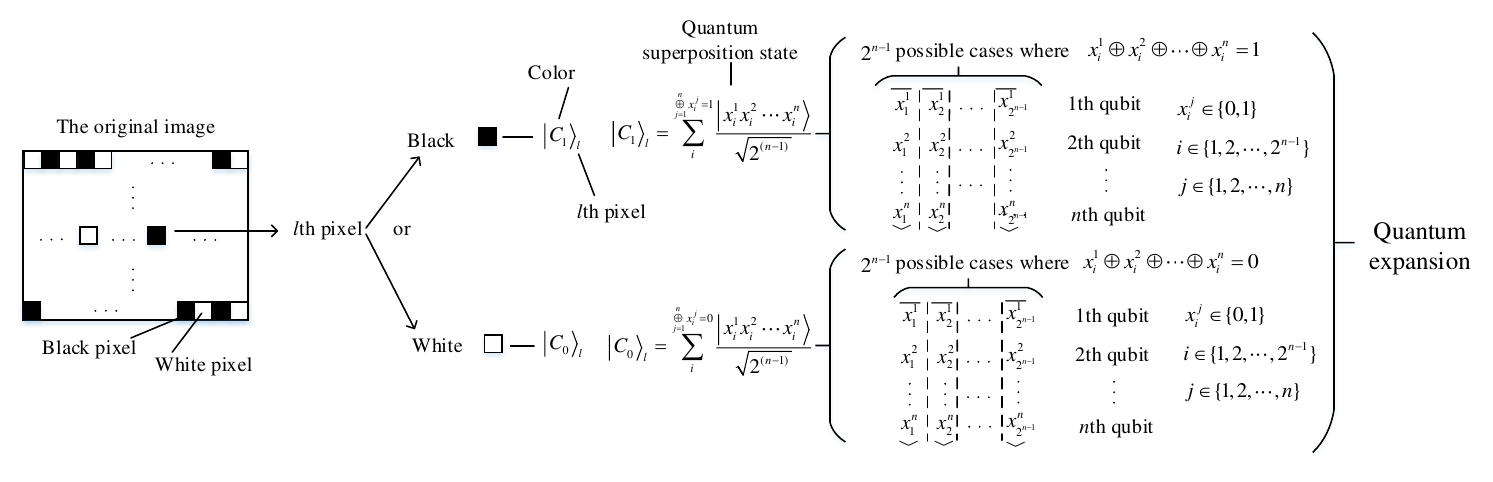}\\
  \caption{The process of encoding color information into quantum superposition state ${\left| {{C_b}} \right\rangle _l}$ ($b=0,1$)}\label{fig4}
\end{figure*}

During the sharing process of ($n$, $n$) QVSS scheme, the most important operation is to encode the color information of each pixel from the secret image, e.g., the $l$th pixel, into an $n$-qubit superposition state ${\left| {{C_b}} \right\rangle _l}$,
\begin{equation}\label{eqn4}
{\left| {{C_b}} \right\rangle _l} = \sum\limits_i^{\mathop  \oplus \limits_{j = 1}^n x_i^j = b} {\frac{{\left| {x_i^1x_i^2 \cdots x_i^n} \right\rangle }}{{\sqrt {{2^{(n - 1)}}} }}},
\end{equation}
where $l \in \left\{ {1,2, \cdots ,s} \right\}$, $i \in \left\{ {1,2, \cdots ,{2^{n - 1}}} \right\}$, $j \in \left\{ {1,2, \cdots ,n} \right\}$, $b \in \left\{ {0,1} \right\}$ and $x_i^j \in \left\{ {0,1} \right\}$. $i$ stands for the $i$th possible case where ${\mathop  \oplus \limits_{j = 1}^n x_i^j = b}$. $j$ represents $j$th qubit and also corresponds to $j$th participant $Bo{b_j}$. The specific process of encoding color information of $l$th pixel into ${\left| {{C_b}} \right\rangle _l}$ is shown in Fig. \ref{fig4}. If the $l$th pixel is white, then $b = 0$ (${\left| {{C_0}} \right\rangle _l}$); if the $l$th pixel is black, then $b = 1$ (${\left| {{C_1}} \right\rangle _l}$). The process of encoding the color information of each pixel into a quantum superposition state can be viewed as quantum expansion, which is analogous to pixel expansion in most classical VSS schemes. Different from the pixel expansion, quantum expansion not only makes the recovered image have no loss in resolution, but also confuses the color information in each share (it makes it impossible for an attacker to directly determine the color information of the pixel).

\begin{figure*}[htbp]
  \centering
  \includegraphics[width=6in]{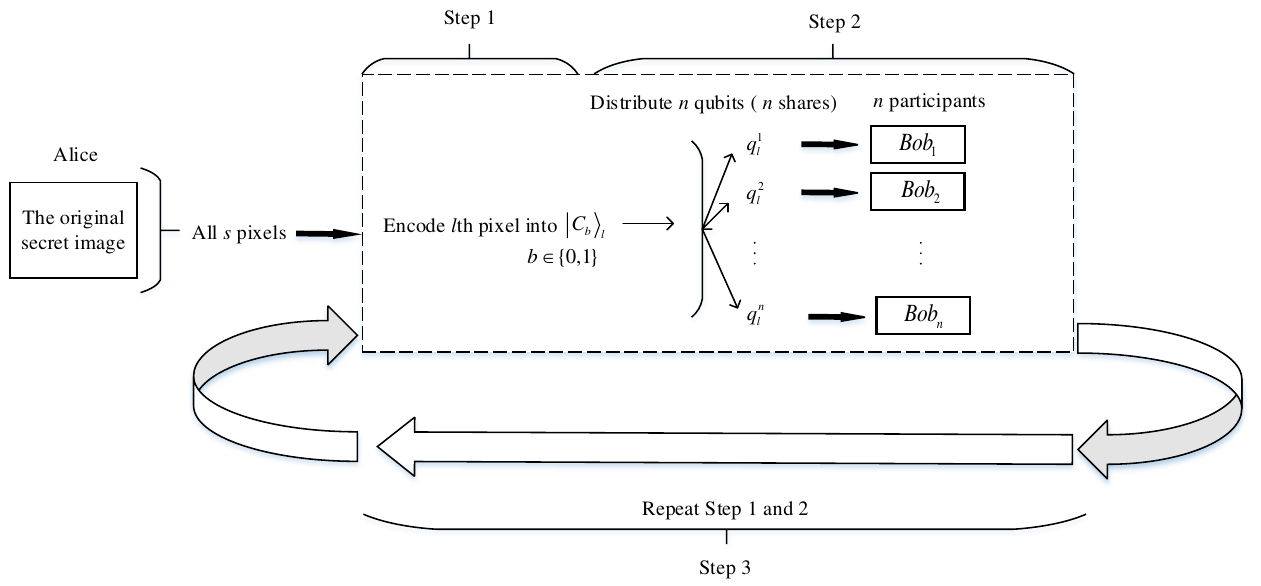}\\
  \caption{The whole sharing process of our proposed QVSS scheme}\label{fig5}
\end{figure*}

The specific steps of the ($n$, $n$) QVSS sharing process are as follows (also shown in Fig. \ref{fig5}).

\noindent \textbf{Step 1}: According to the previous context, Alice encodes the color information of one pixel, e.g., the $l$th pixel, into an $n$-qubit superposition state ${\left| {{C_b}} \right\rangle _l}$.

\noindent \textbf{Step 2}: Alice distributes $n$ qubits $q_l^1$, $q_l^2$, $ \cdots $, $q_l^n$ which compose the state ${\left| {{C_b}} \right\rangle _l}$, as shares to $Bo{b_1}$, $Bo{b_2}$, $ \cdots $, $Bo{b_n}$, respectively.

\noindent \textbf{Step 3}: Alice repeats Step 1 and 2 to handle other $s$-1 pixels.

\subsection{Recovering process}
When all participants want to recover the secret image, we assume that $Bo{b_j}$ is primarily responsible for performing specific operations. Then, he needs to follow the steps below, which is shown in Fig. \ref{fig7}.

\noindent \textbf{Step 1}: $Bo{b_j}$ collects $n$ shares $q_l^1$, $q_l^2$, $ \cdots $, $q_l^n$ which correspond to the $l$th pixel, from remaining $n-1$ participants.

\noindent \textbf{Step 2}: $Bo{b_j}$ selects a measurement base $\left\{ {\left| 0 \right\rangle ,\left| 1 \right\rangle } \right\}$ to measure the $n$-qubit superposition state ${\left| {{C_b}} \right\rangle _l}$ which consists of $q_l^1$, $q_l^2$, $ \cdots $, $q_l^n$. ${\left| {{C_b}} \right\rangle _l}$ collapses into $n$-qubit certain state $\left| {x_i^1x_i^2 \cdots x_i^n} \right\rangle $.

\begin{figure}[htbp]
  \centering
  \includegraphics[width=3in]{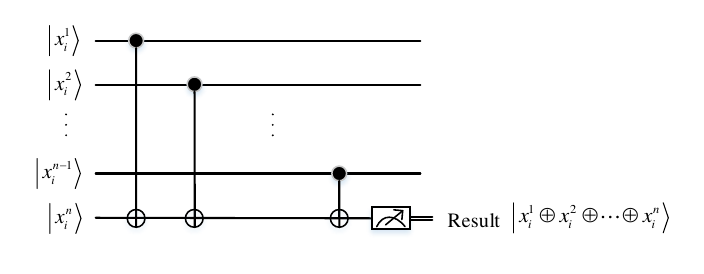}\\
  \caption{The $n$-qubit \textit{XOR} circuit which is composed of multiple \textit{CNOT} quantum gates}\label{fig6}
\end{figure}

\begin{figure*}[htbp]
  \centering
  \includegraphics[width=6in]{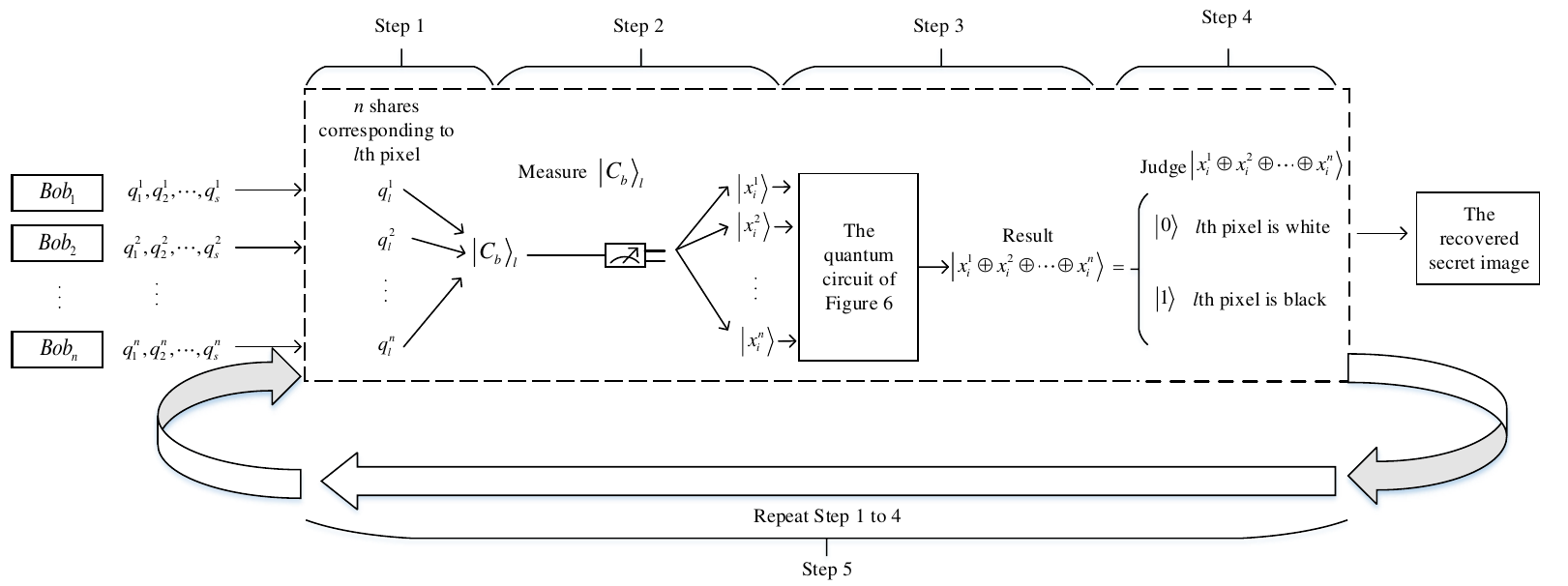}\\
  \caption{The whole recovering process of our proposed QVSS scheme}\label{fig7}
\end{figure*}

\noindent \textbf{Step 3}: $Bo{b_j}$ performs \textit{XOR} operation (its quantum circuit is illustrated in Fig. \ref{fig6}) on the input quantum state $\left| {x_i^1x_i^2 \cdots x_i^n} \right\rangle $ and then get a result state $\left| {x_i^1 \oplus x_i^2 \oplus  \cdots  \oplus x_i^n} \right\rangle $.

\noindent \textbf{Step 4}: $Bo{b_j}$ makes judgment about the result state $\left| {x_i^1 \oplus x_i^2 \oplus  \cdots  \oplus x_i^n} \right\rangle $, if  $\left| {x_i^1 \oplus x_i^2 \oplus  \cdots  \oplus x_i^n} \right\rangle $ $=$ $\left| 0 \right\rangle $ , it means that the $l$th pixel is white, otherwise, the $l$th pixel is black.

\noindent \textbf{Step 5}: $Bo{b_j}$ repeats Step 1 to 4 until the original secret image is recovered.

In our scheme, different from pixel expansion, quantum expansion does not bring about the actual expansion of the shared image size, i.e., the shared image is the same size as the original secret image. So, our scheme does not have loss in resolution and can completely recover the original secret image.

\section{An example: (3, 3) QVSS scheme}
\label{sec:4}
Suppose Alice wants to share a secret image to $Bo{b_1}$, $Bo{b_2}$, $Bo{b_3}$ and the secret image consists of 4 pixels, i.e., $n = 3$ and $s=4$, and the color information of each pixel is shown in Table \ref{tab1}.
\renewcommand\arraystretch{1.5}
\begin{table}[htbp]
\centering
\caption{The color information of each pixel}\label{tab1}
\setlength{\tabcolsep}{7mm}
\begin{tabular}{cc}
\hline
Pixel & Color\\
\hline
1th pixel &  white\\
2th pixel&  black\\
3th pixel & black\\
4th pixel &  white\\
\hline
\end{tabular}
\end{table}

During the sharing process, Alice encodes the color information of each pixel into 3-qubit quantum superposition state ${\left| {{C_b}} \right\rangle _l}$. Then, Alice distributes 3 shares $q_l^1$ (the first qubit), $q_l^2$(the second qubit), and $q_l^3$ (the third one) which compose ${\left| {{C_b}} \right\rangle _l}$, to $Bo{b_1}$, $Bo{b_2}$, $Bo{b_3}$, respectively, where $l \in \{ 1,2,3,4\}$. The quantum superposition states and distributed shares of all pixels in the secret image are listed in Table \ref{tab2}.

\begin{table*}[htbp]
\centering
\caption{The quantum superposition states and distributed shares of all pixel in the sharing process of (3, 3) QVSS scheme}\label{tab2}
\begin{tabular}{ccc}
\hline
Pixel & ${\left| {{C_b}} \right\rangle _l}$  & Shares\\
\hline
1th pixel & ${\left| {{C_0}} \right\rangle _1} = {1 \over 2}(\left| {0_1^10_1^20_1^3} \right\rangle  + \left| {0_2^11_2^21_2^3} \right\rangle  + \left| {1_3^10_3^21_3^3} \right\rangle  + \left| {1_4^11_4^20_4^3} \right\rangle )$ & $q_1^1$, $q_1^2$, $q_1^3$ \\

2th pixel& ${\left| {{C_1}} \right\rangle _2} = {1 \over 2}(\left| {1_1^11_1^21_1^3} \right\rangle  + \left| {0_2^10_2^21_2^3} \right\rangle  + \left| {0_3^11_3^20_3^3} \right\rangle  + \left| {1_4^10_4^20_4^3} \right\rangle )$ & $q_2^1$, $q_2^2$, $q_2^3$ \\

3th pixel & ${\left| {{C_1}} \right\rangle _3} = {1 \over 2}(\left| {1_1^11_1^21_1^3} \right\rangle  + \left| {0_2^10_2^21_2^3} \right\rangle  + \left| {0_3^11_3^20_3^3} \right\rangle  + \left| {1_4^10_4^20_4^3} \right\rangle )$ & $q_3^1$, $q_3^2$, $q_3^3$\\

4th pixel & ${\left| {{C_0}} \right\rangle _4} = {1 \over 2}(\left| {0_1^10_1^20_1^3} \right\rangle  + \left| {0_2^11_2^21_2^3} \right\rangle  + \left| {1_3^10_3^21_3^3} \right\rangle  + \left| {1_4^11_4^20_4^3} \right\rangle )$ & $q_4^1$, $q_4^2$, $q_4^3$ \\
\hline
\end{tabular}
\end{table*}

Suppose $Bo{b_2}$ wants to recover the secret image in the recovering process. In order to recover the first pixel, he should collect all shares $q_1^1$, $q_1^3$ from $Bo{b_1}$ and $Bo{b_3}$. Then, he selects a measurement base $\left\{ {\left| 0 \right\rangle ,\left| 1 \right\rangle } \right\}$ to measure the 3-qubit superposition state ${\left| {{C_0}} \right\rangle _1}$ which is composed of $q_1^1$, $q_1^2$, $q_1^3$. Suppose ${\left| {{C_0}} \right\rangle _1}$ collapses into $\left| {0_1^10_1^20_1^3} \right\rangle $. After that, $Bo{b_2}$ performs \textit{XOR} operation on $\left| {0_1^10_1^20_1^3} \right\rangle $ and get a result state $\left| {0_1^1 \oplus 0_1^2 \oplus 0_1^3} \right\rangle $. So, he can determine that the color of the first pixel is white. Similarly, $Bo{b_2}$ can recover the remaining pixels in the same way, and all the cases for $Bo{b_2}$ are listed in Table \ref{tab3}. By comparing the initial pixel information before the sharing process (see Table \ref{tab1}) and the final pixel information after the recovering process (see the last two columns in Table \ref{tab3}), it can be clearly found that our scheme can completely recover the secret image.

\begin{table*}[htbp]
\centering
\caption{All the cases for $Bo{b_2}$ in the recovering process of (3, 3) QVSS scheme}\label{tab3}
\begin{tabular}{cccccc}
\hline
Shares & ${\left| {{C_b}} \right\rangle _l}$ & Collapsed state & Result state & Color& Pixel\\
\hline
$q_1^1$, $q_1^2$, $q_1^3$ & ${\left| {{C_0}} \right\rangle _1}$ & $\left| {0_1^10_1^20_1^3} \right\rangle $ & $\left| {0_1^1 \oplus 0_1^2 \oplus 0_1^3} \right\rangle  = \left| 0 \right\rangle $ &  white & 1th pixel   \\

$q_2^1$, $q_2^2$, $q_2^3$ & ${\left| {{C_1}} \right\rangle _2}$ & $\left| {1_1^11_1^21_1^3} \right\rangle $ & $\left| {1_1^1 \oplus 1_1^2 \oplus 1_1^3} \right\rangle  = \left| 1 \right\rangle $&  black & 2th pixel \\

$q_3^1$, $q_3^2$, $q_3^3$ & ${\left| {{C_1}} \right\rangle _3}$ & $\left| {1_4^10_4^20_4^3} \right\rangle $ & $\left| {1_4^1 \oplus 0_4^2 \oplus 0_4^3} \right\rangle  = \left| 1 \right\rangle $ & black & 3th pixel  \\

$q_4^1$, $q_4^2$, $q_4^3$ & ${\left| {{C_0}} \right\rangle _4}$ &$\left| {0_2^11_2^21_2^3} \right\rangle $ & $\left| {0_2^1 \oplus 1_2^2 \oplus 1_2^3} \right\rangle  = \left| 0 \right\rangle $ & white &4th pixel  \\
\hline
\end{tabular}
\end{table*}

\section{Correctness analysis}
\label{sec:5}
Here, we mainly prove that the correctness of our scheme, which consists of two criteria: (1) $n$ participants can cooperate to recover the original secret image, (2) less than $n$ participants cannot recover the original one.

\begin{theorem}
The proposed ($n$, $n$) QVSS scheme can recover the original secret image when all the participants work together.
\end{theorem}

\textit{Proof}: Alice encodes each pixel's color information into superposition state ${\left| {{C_b}} \right\rangle _l}$, and distributes $n$ qubits which compose the ${\left| {{C_b}} \right\rangle _l}$ as shares to all participants respectively. When all participants want to recover the original secret image, they need to collect all $n$ shares corresponding to the same pixel and select $Z$ basis to measure them. According to the definition of ${\left| {{C_b}} \right\rangle _l}$, there are ${2^{n - 1}}$ possible cases where make $\left| {x_i^1 \oplus x_i^2 \oplus  \cdots  \oplus x_i^n} \right\rangle  = \left| b \right\rangle $, that is, all possible cases go through Step 3 of the recovering process and only one result state $\left| b \right\rangle $ is obtained. It indicates that the color of the recovered pixel is as same as the original one. So, with $n$ participants involved in the recovering process, the scheme can recover the original secret image.

\begin{theorem}
The proposed QVSS scheme does not work when any $k$ ($k<n$) participants cooperate to recover the original secret image.
\end{theorem}

\textit{Proof}: When $k$ participants want to recover the original secret image, we assume that first $k$ participants collect their $k$ shares to recover the $l$th pixel and $k$ is an even number. After Step 2 of the recovering process, these $k$ shares collapse into $k$-qubit certain state $\left| {x_i^1x_i^2 \cdots x_i^k} \right\rangle $. The state $\left| {x_i^1x_i^2 \cdots x_i^k} \right\rangle $ has ${2^{k}}$ possible cases. Among these ${2^{k}}$ cases, there are $C_k^0 + C_k^2 +  \cdots  + C_k^{k}$ cases that the number of $1$ among $\left| {x_i^1x_i^2 \cdots x_i^k} \right\rangle $ is even, and there are $C_k^1 + C_k^3 +  \cdots  + C_k^{k - 1}$ cases that the number of  $1$ among $\left| {x_i^1x_i^2 \cdots x_i^k} \right\rangle $ is odd. Then, each of ${2^{k}}$ cases goes through Step 3 of recovering process. There are $C_k^0 + C_k^2 +  \cdots  + C_k^{k}$ cases that $\left| {x_i^1 \oplus x_i^2 \oplus  \cdots  \oplus x_i^k} \right\rangle $ $ = $ $\left| 0 \right\rangle $, and there are $C_k^1 + C_k^3 +  \cdots  + C_k^{k-1}$ cases that $\left| {x_i^1 \oplus x_i^2 \oplus  \cdots  \oplus x_i^k} \right\rangle $ $ = $ $\left| 1 \right\rangle $. According to the nature of combination number, $C_k^0 + C_k^2 +  \cdots  + C_k^{k} = C_k^1 + C_k^3 +  \cdots  + C_k^{k-1}$. So, the probabilities of getting $\left| 0 \right\rangle $ and $\left| 1 \right\rangle $ are both equal to $\frac{1}{2}$. Similarly, when $k$ is an odd number, the probabilities of getting $\left| 0 \right\rangle $ and $\left| 1 \right\rangle $ are also equal to $\frac{1}{2}$. We can see that it is impossible to determine whether the $l$th pixel is white or black.

Overall, we can see that with respecting the two criteria, the proposed scheme is a well-defined ($n$, $n$) QVSS scheme.

\section{Comparison and Discussion}
\label{sec:6}
In order to evaluate our scheme, we chose a classical VSS scheme \cite{Naor95} and two quantum VSS schemes \cite{Song14,Das15} as references, and compare our QVSS scheme with them from the following aspects: single-pixel parallel processing, pixel expansion, and the loss in resolution. For a more intuitive representation, the results of the comparison are shown in Table \ref{tab4}.

As mentioned in Sect. \label{sec:2}, the pixel is treated as a separate basic unit throughout Naor \textit{et al.}'s scheme. So, this scheme has the advantage of single-pixel parallel processing. In order to make individual participant get no information from shares, the scheme use the strategy of pixel expansion to confuse the information of subpixels in each share. However, the size of the shared image is $m$ times the original secret image and cause the loss in resolution from the original image to the recovered one.

Then, we compare with two quantum VSS schemes, Song \textit{et al.}'s scheme \cite{Song14} and Das \textit{et al.}'s scheme \cite{Das15}. In the Song \textit{et al.}'s scheme, the whole secret image is repeatedly encoded into a quantum state. So, the scheme does not retain the advantage of single-pixel parallel processing, i.e., it is not suitable for parallel computing of images. But, the scheme splits the original secret image into sub-images as shares with multiple measurement operations. Therefore, it does not need the strategy of pixel expansion. And the restored image is exactly the same as the original secret image. Different from Song \textit{et al.}'s scheme, one pixel is treated as a unit for parallel processing in the Das \textit{et al.}'s scheme. However, the scheme still uses the strategy of pixel expansion and apply the characteristics of quantum mechanics to determine the color of each sub-pixel in each share, where each share has multiple sub-pixels. So, the recovered image has the loss in resolution in the Das \textit{et al.}'s scheme.

As same as Naor \textit{et al.}'s scheme and Das \textit{et al.}'s scheme \cite{Das15}, our scheme retains the advantage of single-pixel parallel processing, i.e., the color information of each pixel is encoded into $n$-qubit superposition state. Different from them in other aspects, we use the strategy of quantum expansion instead of pixel expansion. Therefore, the size of the shared image is as same as the original one. In the recovering process, all participants cooperate to measure the qubits they hold and execute the $n$-qubit \textit{XOR} operation to recover each pixel of the secret image. The recovered image is the same as the original secret image and has no loss in resolution.

\begin{table*}[htbp]
\centering
\caption{Comparison among analogous quantum schemes, Naor \textit{et al.}'s scheme and our scheme}\label{tab4}
\setlength{\tabcolsep}{4mm}
\begin{tabular}{ccccc}
\hline
Schemes & Naor \textit{et al.}'s scheme\cite{Naor95} & Song \textit{et al.}'s scheme\cite{Song14}  &Das \textit{et al.}'s scheme\cite{Das15}   &Our scheme \\
\hline
Single-pixel parallel processing  &  Yes &No &Yes & Yes\\
Pixel expansion                        &  Yes &No &Yes& No\\
The loss in resolution               & Yes  &No &No&No\\
\hline
\end{tabular}
\end{table*}

\section{Experiment implementation with IBM Q}
\label{sec:7}
IBM Q \cite{IBM18} is an online platform that gives users in the general public access to a set of IBM's prototype quantum processors via the network. In this section, we use IBM Q to demonstrate the practical feasibility of this scheme.

For the sake of brevity, we assume that $n=6$, i.e., the (6, 6) QVSS scheme is the experimental object. In Step 1 of the sharing process, to share a white pixel, we encode color information of the pixel into the 6-qubit superposition state $\left| {{C_0}} \right\rangle $ which is composed of 6 qubits. $\left| {{C_0}} \right\rangle $ is composed of the base states in Table \ref{tab5}, and each base state in the quantum superposition state $\left| {{C_0}} \right\rangle $ is with equal probability.

\begin{equation}\label{eqn5}
\left| {{C_0}} \right\rangle  = \sum\limits_i^{\mathop  \oplus \limits_{j = 1}^6 x_i^j = 0} {{{\left| {x_i^1x_i^2 \cdots x_i^6} \right\rangle } \over {\sqrt {32} }}}
\end{equation}

\begin{figure*}[htbp]
  \centering
  \includegraphics[width=5in]{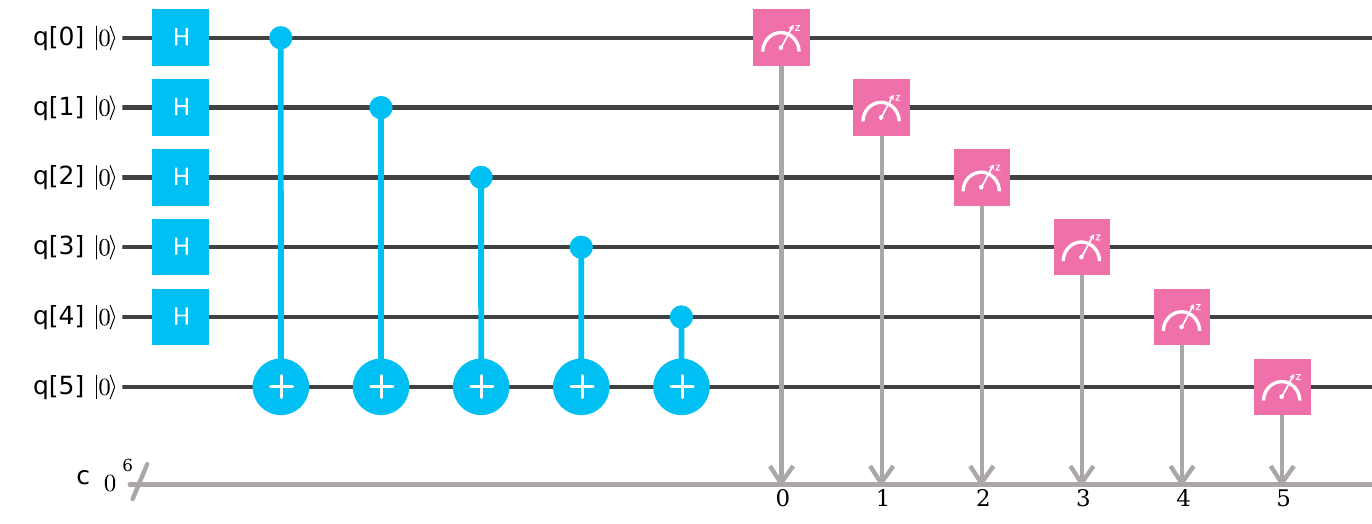}\\
  \caption{Implementation circuit of preparing the quantum superposition state $\left| {{C_0}} \right\rangle $}\label{fig8}
\end{figure*}

\begin{figure*}[htbp]
  \centering
  \includegraphics[width=6in]{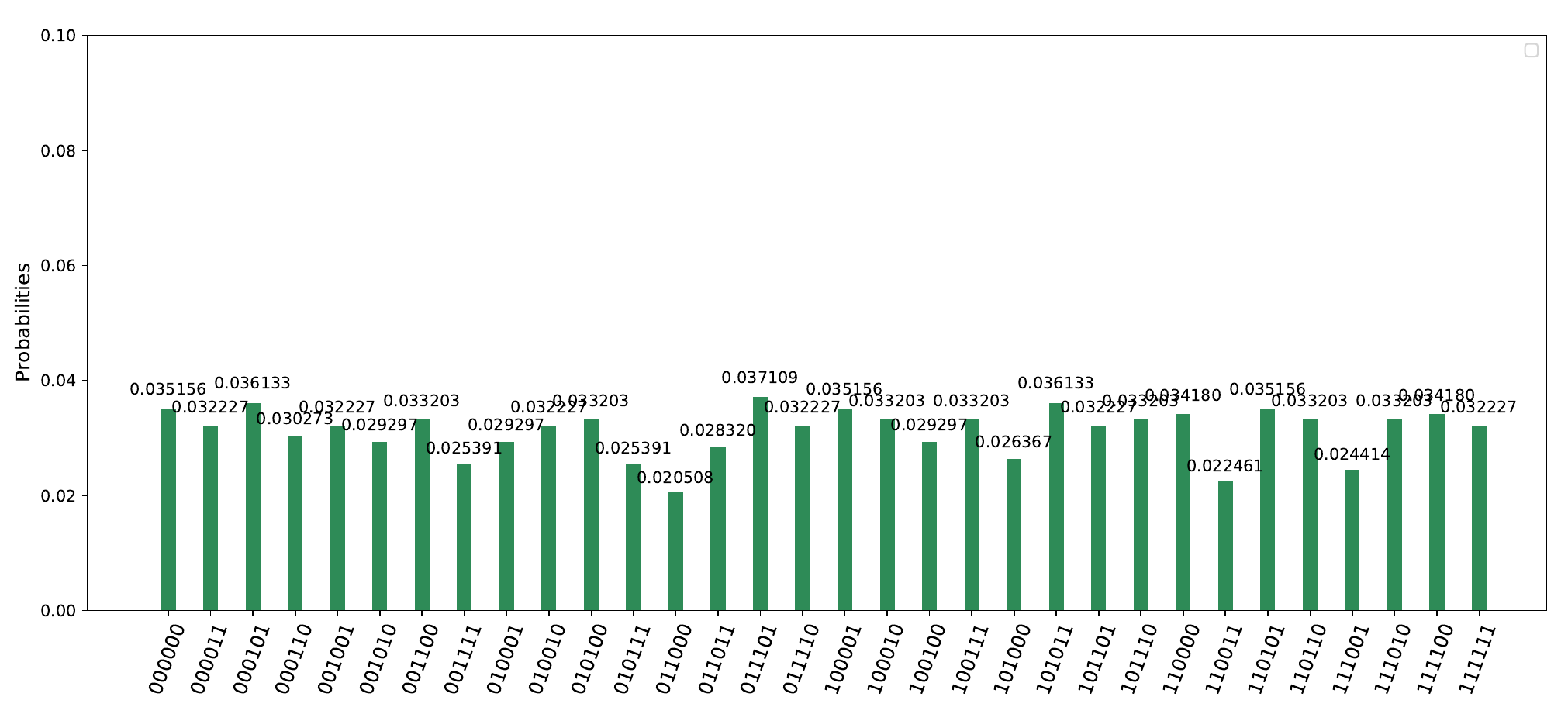}\\
  \caption{The probability of each base state in the quantum superposition state $\left| {{C_0}} \right\rangle $}\label{fig9}
\end{figure*}

Since IBM Q is just a single quantum computer, which does not support the transfer of quantum information between multiple nodes, so we ignore the process of qubits transmission in our scheme (i.e., Step 2 of the sharing process and Step 1 of the recovering process) in our quantum experiment, and directly implement the reminder in IBM Q. In the sharing process, we firstly design a quantum circuit (shown in Fig. \ref{fig8}) to construct $\left| {{C_0}} \right\rangle $, and run it on the IBM Q platform. We measure the state $\left| {{C_0}} \right\rangle $ and then get the probability of each base state in $\left| {{C_0}} \right\rangle $, where the result is shown in Fig. \ref{fig9}.  We can find that the probability of each state is almost equal, and the probability amplitudes of all base states float above and below $\frac{1}{{\sqrt {32} }}$.

\begin{table*}[htbp]
\centering
\caption{All base states of the quantum superposition state $\left| {{C_0}} \right\rangle $ }\label{tab5}
\begin{tabular}{cccccccc}
\hline
Number & Base state & Number & Base state & Number & Base state & Number & Base state \\
\hline
1 &  $\left| {000000} \right\rangle$  &  9 &  $\left| {010001} \right\rangle$  &  17 &  $\left| {100001} \right\rangle$  &  25 &  $\left| {101110} \right\rangle$  \\
2& $\left| {000011} \right\rangle$  &  10&  $\left| {010010} \right\rangle$  &  18&  $\left| {100010} \right\rangle$ &  26&  $\left| {110011} \right\rangle$  \\
3 & $\left| {000101} \right\rangle$  &  11 &$\left| {010100} \right\rangle$  &  19 &  $\left| {100100} \right\rangle$  &  27&  $\left| {110101} \right\rangle$ \\
4& $\left| {000110} \right\rangle$  &  12 &  $\left| {010111} \right\rangle$  &  20 &  $\left| {100111} \right\rangle$  & 28 &  $\left| {110110} \right\rangle$ \\
5 &  $\left| {001001} \right\rangle$  &  13 &  $\left| {011000} \right\rangle$  & 21 &  $\left| {101000} \right\rangle$  &  29 &  $\left| {111001} \right\rangle$ \\
6&  $\left| {001010} \right\rangle$  &  14&  $\left| {011011} \right\rangle$  &  22&  $\left| {101011} \right\rangle$  &  30&  $\left| {111010} \right\rangle$ \\
7 & $\left| {001100} \right\rangle$  &  15 &  $\left| {011101} \right\rangle$ &  23 & $\left| {101101} \right\rangle$  &  31&  $\left| {111100} \right\rangle$  \\
8&  $\left| {001111} \right\rangle$   &  16 &  $\left| {011110} \right\rangle$  &  24 &  $\left| {101110} \right\rangle$  & 32 &  $\left| {111111} \right\rangle$ \\
\hline
\end{tabular}
\end{table*}

\begin{figure*}[htbp]
\centering
\subfigure[Implementation circuit of \textit{XOR} operation on $\left| {101000} \right\rangle $ ]{
\includegraphics[width=3in]{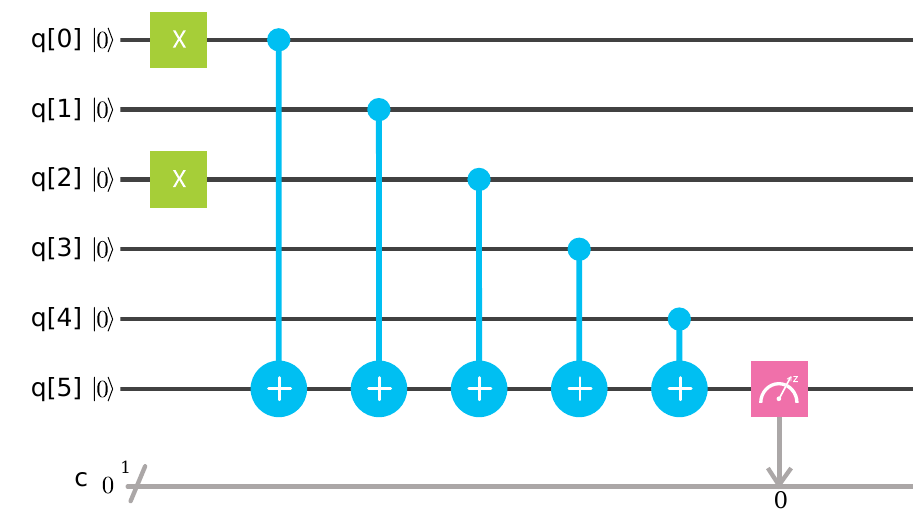}
}
\quad
\subfigure[The probability of the result state]{
\includegraphics[width=3in]{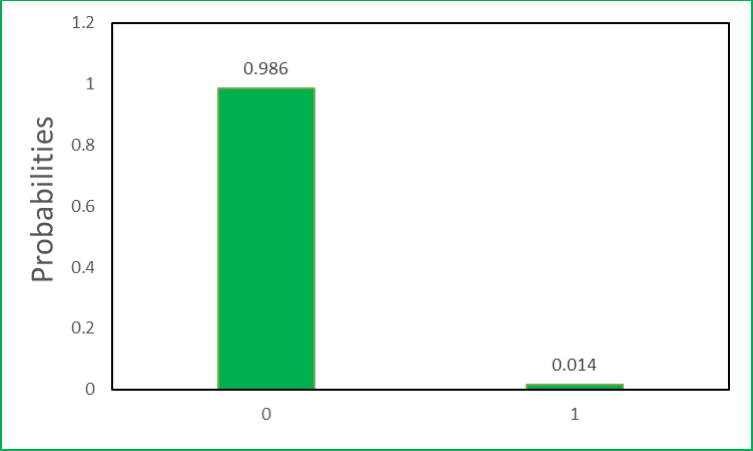}
}
\caption{ The process of performing \textit{XOR} operation on $\left| {101000} \right\rangle $}
\label{fig10}
\end{figure*}

After the state $\left| {{C_0}} \right\rangle $ is measured, the quantum superposition state may collapse into one certain state from 32 base states. We assume that $\left| {{C_0}} \right\rangle $ collapses into one certain state $\left| {101000} \right\rangle $. Then, we perform \textit{XOR} operation on the state $\left| {101000} \right\rangle $ to get a result state on the IBM Q, where the circuit is shown in Fig. \ref{fig10} (a). The result state is $\left| 0 \right\rangle $ and its probability is shown in Fig. \ref{fig10} (b). Finally, according to Step 4 of the recovering process, we can determine that the pixel is white.

To share a black pixel, the process of experiment is similar to the above. We encode color information of the pixel into the quantum superposition state $\left| {{C_1}} \right\rangle $. $\left| {{C_1}} \right\rangle $ can be composed of each base state in Table \ref{tab6}.  Each base state in the quantum superposition state $\left| {{C_1}} \right\rangle $ is also with equal probability, so

\begin{equation}\label{eqn6}
\left| {{C_1}} \right\rangle  = \sum\limits_i^{\mathop  \oplus \limits_{j = 1}^6 x_i^j = 1} {{{\left| {x_i^1x_i^2 \cdots x_i^6} \right\rangle } \over {\sqrt {32} }}}.
\end{equation}

\begin{table*}[htbp]
\centering
\caption{All base states of the quantum superposition state $\left| {{C_1}} \right\rangle $ }\label{tab6}
\begin{tabular}{cccccccc}
\hline
Number & Base state & Number & Base state & Number & Base state & Number & Base state \\
\hline
1 &  $\left| {000001} \right\rangle$  &  9 &  $\left| {010000} \right\rangle$  &  17 &  $\left| {100000} \right\rangle$  &  25 &  $\left| {101111} \right\rangle$  \\
2& $\left| {000010} \right\rangle$  &  10&  $\left| {010011} \right\rangle$  &  18&  $\left| {100011} \right\rangle$ &  26&  $\left| {110010} \right\rangle$  \\
3 & $\left| {000100} \right\rangle$  &  11 &$\left| {010101} \right\rangle$  &  19 &  $\left| {100101} \right\rangle$  &  27&  $\left| {110100} \right\rangle$ \\
4& $\left| {000111} \right\rangle$  &  12 &  $\left| {010110} \right\rangle$  &  20 &  $\left| {100110} \right\rangle$  & 28 &  $\left| {110111} \right\rangle$ \\
5 &  $\left| {001000} \right\rangle$  &  13 &  $\left| {011001} \right\rangle$  & 21 &  $\left| {101001} \right\rangle$  &  29 &  $\left| {111000} \right\rangle$ \\
6&  $\left| {001011} \right\rangle$  &  14&  $\left| {011010} \right\rangle$  &  22&  $\left| {101010} \right\rangle$  &  30&  $\left| {111011} \right\rangle$ \\
7 & $\left| {001101} \right\rangle$  &  15 &  $\left| {011100} \right\rangle$ &  23 & $\left| {101100} \right\rangle$  &  31&  $\left| {111101} \right\rangle$  \\
8&  $\left| {001110} \right\rangle$   &  16 &  $\left| {011111} \right\rangle$  &  24 &  $\left| {101111} \right\rangle$  & 32 &  $\left| {111110} \right\rangle$ \\
\hline
\end{tabular}
\end{table*}

\begin{figure*}[htbp]
  \centering
  \includegraphics[width=5in]{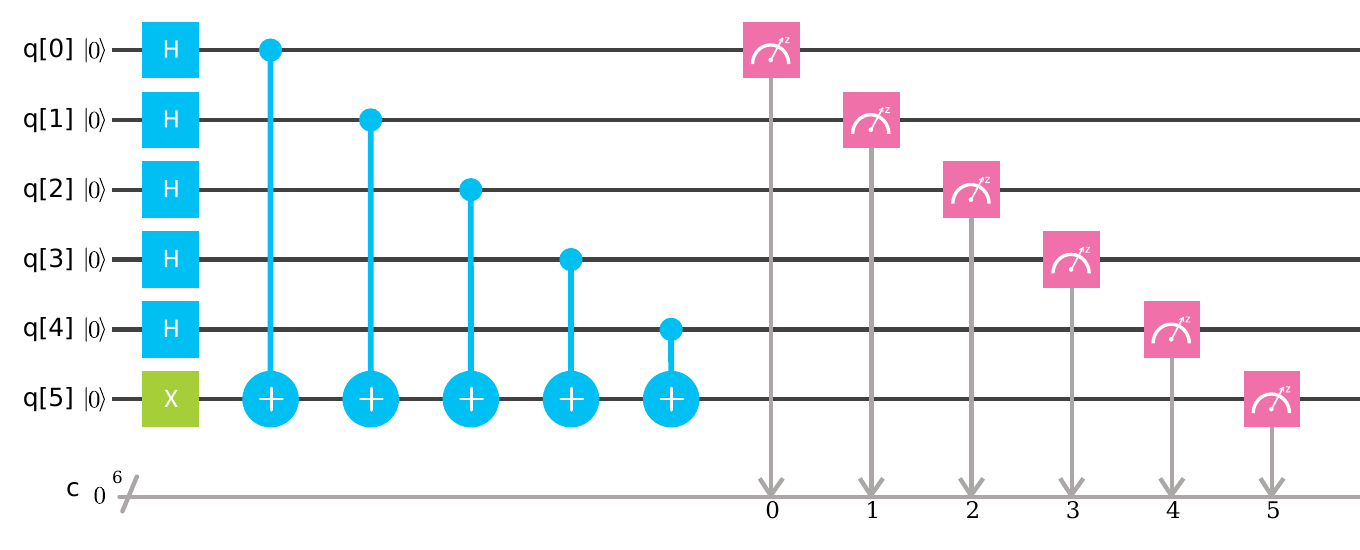}\\
  \caption{Implementation circuit of preparing the quantum superposition state $\left| {{C_1}} \right\rangle $}\label{fig11}
\end{figure*}

\begin{figure*}[htbp]
  \centering
  \includegraphics[width=6in]{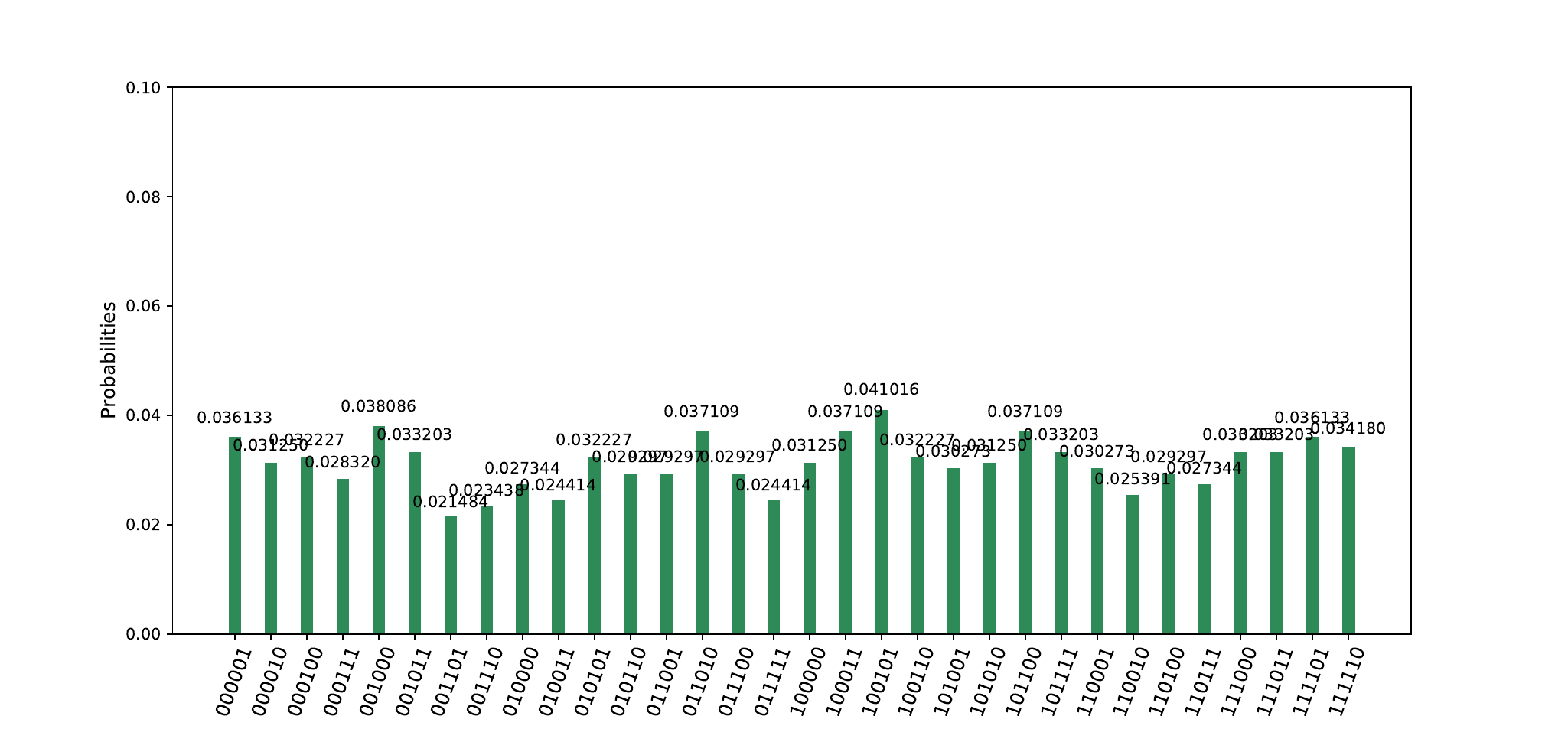}\\
  \caption{The probability of each base state in the quantum superposition state $\left| {{C_1}} \right\rangle $}\label{fig12}
\end{figure*}

\begin{figure*}[htbp]
\centering
\subfigure[Implementation circuit of \textit{XOR} operation on $\left| {101010} \right\rangle $ ]{
\includegraphics[width=3in]{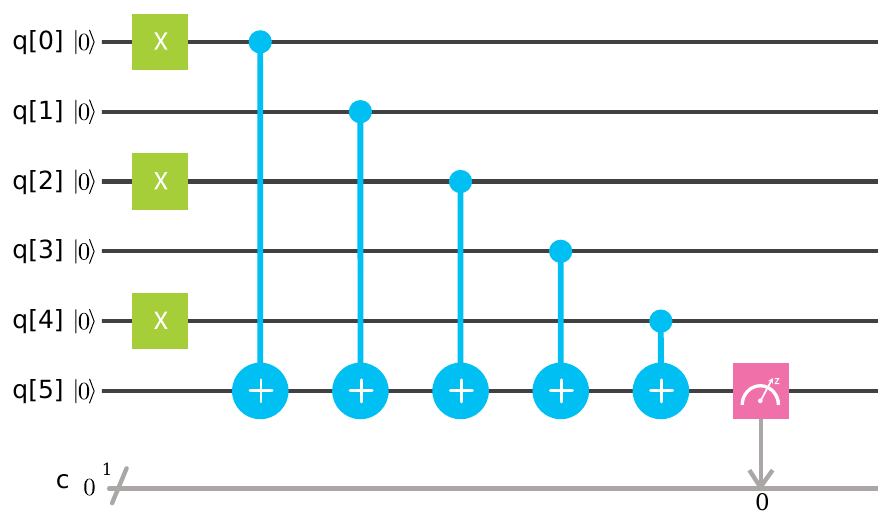}
}
\quad
\subfigure[The probability of the result state]{
\includegraphics[width=3in]{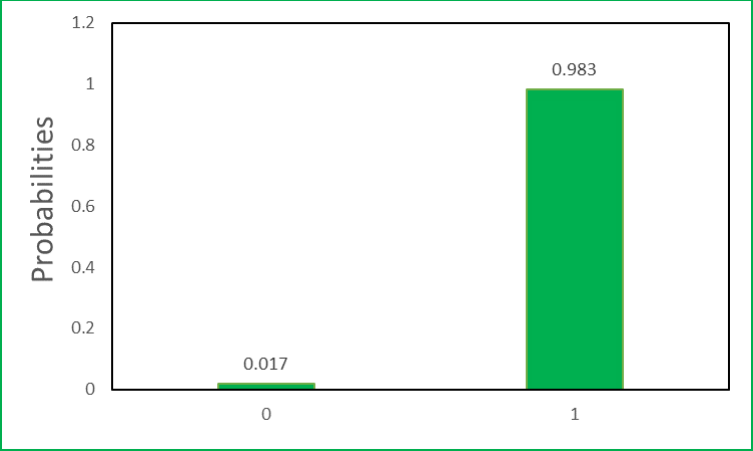}
}
\caption{The process of  performing \textit{XOR} operation on $\left| {101010} \right\rangle $}
\label{fig13}
\end{figure*}

The process of constructing $\left| {{C_1}} \right\rangle $ and the probability of each base state in $\left| {{C_1}} \right\rangle $ are shown in Fig. \ref{fig11} and Fig. \ref{fig12}, respectively. After the state $\left| {{C_1}} \right\rangle $ is measured, we assume that $\left| {{C_1}} \right\rangle $ collapses into one certain state $\left| {101010} \right\rangle $. The process of \textit{XOR} operation and the probability of  the result state are shown in Fig. \ref{fig13} (a) and (b) respectively. Finally, according to Step 4 of the recovering process, we can determine that the pixel is black.

\section{Conclusion}
\label{sec:8}

This paper presents a novel ($n$, $n$) quantum visual secret sharing scheme based on Naor \textit{et al.}'s scheme. In our scheme, we encode the color of each pixel into $n$-qubit superposition state, to preserve the advantage of single-pixel parallel processing. Moreover, our scheme uses the strategy of quantum expansion instead of classical pixel expansion. This strategy solves the problem that the recovered image has loss in resolution which caused by pixel expansion. Moreover, compared with other classical VSS scheme and two quantum VSS schemes, we can see that our scheme preserves the advantage of single-pixel parallel processing and solves two problems, which are pixel expansion and the loss in resolution. Besides, the proposed scheme is able to meet two criteria (shown in Sect. \ref{sec:5}) and has practical feasibility. Of course, ($t$, $n$) scheme is a more application-oriented scheme. How to construct a threshold scheme is one of our future work.

\section*{Acknowledgment}

The authors would like to express heartfelt gratitude to the anonymous reviewers and editor for their comments that improved the quality of this paper. And the support of all the members of the quantum research group of NUIST is especially acknowledged, their professional discussions and advice have helped us a lot.

\end{document}